\documentclass[letterpaper, 10 pt, conference]{ieeeconf}  

\IEEEoverridecommandlockouts                              
\usepackage[latin1]{inputenc}

\usepackage{tikz}
\newcommand{\cod}{0}
\newcommand{\co}{170}
\definecolor{pink}{RGB}{\co, \co, \co}
\definecolor{lpink}{RGB}{\cod, \cod, \cod}
\definecolor{blue}{RGB}{\cod,\cod,\cod}
\definecolor{lblue}{RGB}{\cod,\cod,\cod}
\definecolor{dgreen}{RGB}{40,120,40}
\definecolor{dblue}{RGB}{30,30,200}

\usepackage{xstring}
\usepackage{etoolbox}
\usepackage{refcount}
\usepackage{amsmath}

\usepackage{amsthm}
\newtheorem{ex}{Example}
\newenvironment{exbis}[1]
  {
   \addtocounter{ex}{-1}%
   \begin{ex} } 
  {\end{ex}}

\newtheorem{thm}{Theorem}
\newtheorem{defi}{Definition}
\newtheorem{prop}{Proposition}

\newtheorem{lem}{Lemma}

\newtheorem{prob}{Problem}

\newtheorem{conj}{Conjecture}

\usepackage{cite}

\usepackage{color}
\usepackage{amsfonts}
\usepackage{amssymb}
\usepackage{graphicx}
\usepackage{algorithmicx}
\usepackage{algpseudocode}
\usepackage{color}
\usepackage{amssymb}

\title{\LARGE \bf
Reachability of Consensus and Synchronizing Automata}

\author{Pierre-Yves Chevalier, Julien M. Hendrickx, Raphaël M. Jungers%
\thanks{All authors are members of ICTEAM, Universit\'e catholique de Louvain, Belgium.  Their work is supported by the Belgian Network DYSCO, funded by the Belgian government and the Concerted Research Action (ARC) of the French Community
of Belgium. R. M. Jungers is an F.R.S.-FNRS research associate. Contact: {\tt \{pierre-yves.chevalier, julien.hendrickx, raphael.jungers\}@uclouvain.be}.
 The authors would also like to thank François Gonze for helpful discussions.}
}

\begin{document}

\maketitle
\thispagestyle{empty}
\pagestyle{empty}

\begin{abstract}
We consider the problem of determining the existence of a sequence of matrices driving a discrete-time multi-agent consensus system to consensus. 
We transform this problem into  the problem  of the existence of a product of the   (stochastic) transition matrices   that has a positive column.  This allows us to make use of results from automata theory to sets of stochastic matrices.  Our  main result  is  a polynomial-time algorithm to decide the existence of a sequence of matrices achieving consensus.
\end{abstract}

\section{Introduction}

Consensus systems represent groups of agents trying to reach agreement on some value.
They are commonly used in many distributed computation systems, and have attracted much research attention in recent years. 
 Indeed, many decentralized systems are a combination of local computing and global synchronization, and consensus systems are an appropriate tool to perform the synchronization step.
The simplest consensus system consists of agents computing  the  weighted average of values of other agents: $$x(t+1) = A_t x(t),$$ with $A_t$ \emph{stochastic matrices}, i.e., their entries are nonnegative and the entries on each row sum up to one.  

Recent works have considered the problem of controlling consensus systems. 
This research deals, for instance, with finding conditions on $A$ and $B$ under which system $$x(t+1) = Ax(t) + Bu$$  can be steered into any desired configuration \cite{rahmani2, chapman}. 

We consider a different kind of controllability: the system is not controlled by an exogenous input, but by choosing the matrix of interaction $A_t$ at each time.
 Consider, for instance, a wireless network of agents trying to converge to consensus.  One solution to avoid interference is to partition the agents into groups which emit at different times.  Thus, the problem arises of optimally scheduling the communication protocol. 
 The problem consists here in finding a scheduling such  that the agents converge to consensus. 

Formally, we study the system 
\begin{equation} \begin{aligned}x(t+1) &= A_{\sigma(t)}x(t) \\
x(0) &= x_0,
\end{aligned} \label{sys}
\end{equation} and we want to solve the following decision problem.
\begin{prob} Given a set of stochastic matrices $ M  = \{A_1, \dots , A_m\}$, does there exist a  switching  sequence $\sigma: \mathbb{N} \mapsto \{1, \dots, m\} : t \mapsto \sigma(t)$ such that, for any $x_0$, System (\ref{sys}) converges to consensus, i.e., to  some vector $a \textbf{1} = a \begin{pmatrix} 1 & \dots & 1 \end{pmatrix}^\top$ for some $a$?\footnote{This problem is Problem 2 of our article \cite{CHJ12}, that we restrict here to the case of stochastic matrices.  We obtain here much stronger results with a different approach.}
\label{pb21}
\end{prob}

This problem can be seen as an \emph{open loop} control problem. Indeed, it deals with the existence of a sequence $\sigma$ that steers System (\ref{sys}) to consensus \emph{from any initial condition}. 
One could have $\sigma$ depend on the initial condition $x_0$ or, more generally, on the state $x(t)$ (closed loop control), but we proved in \cite[Proposition 1.b]{CHJ12}  that the two are equivalent.
 

Our problem can be seen as deciding stabilizability of a switched system with control on the switching signal.  This problem has motivated much research effort (see, e.g. \cite{geromel2006stability, stanford} and \cite[Section 2.2.4]{jungers_lncis}) and is known to be very hard.  For instance, 
 deciding, for a matrix set $ M $, whether there is a product of matrices from $ M $ that converges to zero is an undecidable problem \cite{paterson}. 


We will show that for consensus systems, not only is the problem decidable, but it is decidable in polynomial time. Our proof technique proceeds in two steps. First, we reduce the problem to that of determining  the existence of  a \emph{positive-column  product}, i.e., a product of transition matrices that has a positive column. 
We call a \emph{ positive-column  word} the sequence of indices of a positive-column  product. 
It turns out that the existence of positive-column  words has been extensively studied for sets of \emph{binary} stochastic matrices (i.e., stochastic matrices with the additional constraint that the entries are in $\{0,1\}$). Sets of binary stochastic matrices that have a positive-column  word are called \emph{synchronizing semi-automata}. 
Secondly, we leverage results on synchronizing automata and extend them to sets of stochastic matrices. 

\subsection{Synchronizing Automata}

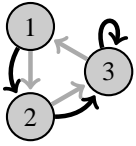
\begin{figure}[h!]
\centering
\usetikzlibrary{arrows}
\newcommand{\ec}{1.2}
\newcommand{\ba}{.7}
\newcommand{\wi}{.6}
\begin{tikzpicture}[->,shorten >=1pt,auto,node distance=2cm,
  thick,main node/.style={circle,fill=blue!20,draw}]

  \node[main node] (1) {1};
  \node[main node] (2) at (0, -\ec) {2};
  \node[main node] (3) at (.866*\ec, -\ec/2) {3};
  

  \path[every node/.style={font=\sffamily\small}]
   (1) edge [draw=blue, bend right, line width=\wi mm] node [left] {} (2)
   (1) edge [draw=pink, line width=\wi mm] node [left] {} (2)
   (2) edge [draw=blue, bend right, line width=\wi mm] node [left] {} (3)
   (2) edge [draw=pink, line width=\wi mm] node [left] {} (3)
   (3) edge [draw=pink, line width=\wi mm] node [left] {} (1)
   
%
%

   (3) edge [draw=blue, loop above, distance=4*\ec mm, line width=\wi mm] node [left] {} (3)

;
\end{tikzpicture}
\caption{A simple semi-automaton with three states and two actions (black and grey edges). } 
\label{f4}
\end{figure}

Synchronizing automata appeared in theoretical computer science in the sixties and have attracted lots of research attention. 
 A semi-automaton is a triple $(S, \Sigma, M)$
where $S$ is a finite set of \emph{states},  $\Sigma $  is a finite set of \emph{letters} and is called the \emph{alphabet}, and $M$ is a finite set of mappings from $S$ to $S$. 
An automaton is a semi-automaton with a set of ACCEPT states. These ACCEPT states determine which input words are accepted by the automaton.  
 Semi-automata can be represented as sets of digraphs  where each node has outdegree one. An example is given  in  Figure \ref{f4}.
They can also be represented 
by a set of matrices, containing the adjacency matrices  of these digraphs.
 Since the transition function is deterministic, each node of each digraph has outdegree one,  and the adjacency matrices of the digraphs    have a  single  one on each row and zero everywhere else. Hence, they are binary and stochastic. The set corresponding to the semi-automaton of Figure \ref{f4} is $$ M = \left\{A_1 = \begin{pmatrix} 0&1&0\\0&0&1\\1&0&0\end{pmatrix}, A_2 = \begin{pmatrix} 0&1&0\\0&0&1\\0&0&1\end{pmatrix}\right\}.$$

 Automata and semi-automata were initially introduced as simple models of computing devices: the actions representing different control commands, and the states representing different possible states of the memory registers of the device.  A semi-automaton is said to be synchronizing if there is a sequence of actions, independent of the state, that reinitializes the device, i.e., that drives the memory onto a particular state. This sequence of actions is called a \emph{synchronizing word}. 
 Automata and  synchronizing automata have generated intense research efforts in theoretical computer science \cite{Moore56, AdlerWeiss70, trahtman, Pin83a, volkov_survey, GJ15}, and provided several deep results.

Recently, several works made connections between synchronizing automata and topics in systems and control.  
For instance, \cite{jungers_sync_12} applied convex optimization tools to the study of synchronizing automata and \cite{fominykh} linked synchronization with game theoretical concepts.  

\subsection{Equivalent Problems and Applications}
The question of convergence can also be asked in the case where the matrices come each with a certain probability at every time. 
We will see (Proposition \ref{eq_random}) that a positive answer to Problem \ref{pb21} is equivalent to convergence to consensus 
 with probability 1 
 when the switching is random, each matrix is chosen with nonzero probability and choices of matrices at different times are independent\footnote{In fact, if the choices of matrices are not independent, the equivalence still holds provided that each matrix is chosen with nonzero \emph{conditional} probability.}. 
Therefore, our investigation of Problem \ref{pb21} yields results that are also applicable to consensus systems with random switching.

Problem \ref{pb21} is also equivalent to that of determining whether an inhomogeneous Markov chain may be mixing, i.e., may forget its initial condition. Indeed, an inhomogeneous Markov chain can be represented as the transpose of System (\ref{sys}). 
 Moreover, thanks to Proposition \ref{eq_random}, Problem \ref{pb21} is also equivalent to the question of whether an inhomogeneous markov chain is mixing with probability 1. More generally, our results apply to any process with column-stochastic transition matrices, i.e., matrices whose transposes are stochastic\footnote{Stochastic matrices are sometimes referred to as \emph{row-stochastic}.}. In fact, any positive linear process that preserves the sum of the elements of the state vector can be represented with column-stochastic transition matrices. 
A particular case is the push-sum algorithm, a decentralized method to compute an average \cite{kempe, benezit}. 
In this algorithm, agents have initial values $x_i(0)$ and they want to compute their average. Each agent $i$ has two values  $s_i(t)$ and $w_i(t)$. $s_i(t)$ is a fraction of the sum of $x_i(0)$s and $w_i(t)$ is a weight value. Each agent's estimate of the average is $x(t) = s(t)/w(t)$. 
At times, an agent sends a fraction of its values $s_i(t), w_i(t)$ to another agent. Under suitable assumptions, the ratio $s_i(t)/w_i(t)$ converges to the average of the $x_i(0)$s.
This process  preserves the sum of the values of agents and can therefore be represented with column-stochastic transition matrices.
Hence, the results presented in this article automatically apply to the convergence analysis of the push-sum algorithm.

\subsection{Outline}
 
In Section \ref{s:classical}, we present some results on automata that we generalize later in the article. 
In Section \ref{s:from_to}, we prove the equivalence between Problem \ref{pb21} and \emph{column-primitivity} of the set $ M $, i.e., the existence of a positive-column  product. 
In Sections \ref{s:results} and \ref{ss:np-hard}, we extend different results known for automata to finite sets of \emph{nonnegative matrices with no zero row} (of which stochastic matrices are a subset). 
We obtain 
\renewcommand\labelitemi{\tiny$\bullet$}
\begin{itemize}
\item a polynomial bound ($O(n^3)$) on the length of positive-column words (Theorem \ref{pin_stoch}),  
\item a polynomial-time ($O(n^4)$) algorithm to decide the existence of a positive-column  word (Theorem \ref{pairs_algo_bis}), 
\item a proof of NP-hardness of finding the shortest positive-column  word, which also holds for sets of matrices with positive diagonals 
(Theorem \ref{np-hard_thm}). 
\end{itemize}

\section{Classical Results on Automata}
\label{s:classical}

We state in this section a couple of classical results on automata. We will extend them to finite sets of nonnegative matrices with no zero row in order to obtain results on consensus systems. We recall that an automaton is said to be synchronizing if it has a positive-column  product.

\begin{conj}[{\v C}ern{\'y} Conjecture \cite{Cerny64}] If a semi-automaton is synchronizing then it has a synchronizing word of length at most $(n-1)^2$.\label{cerny_conj}
\end{conj}

The best proven bound, however, is given by the next theorem.

\begin{thm}[Pin, Frankl \cite{Pin83a, Frankl82}]If a semi-automaton is synchronizing then it has a synchronizing word of length at most $\frac{n^3-n}{6}$.\label{cerny_thm}
\end{thm}

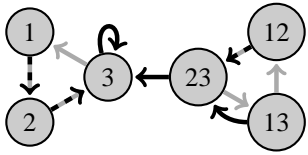
\begin{figure}[h!]
\centering
\usetikzlibrary{arrows}
\newcommand{\ec}{1.2}
\newcommand{\hauteur}{.866}
\newcommand{\ba}{.7}
\newcommand{\wi}{.6}
\begin{tikzpicture}[->,shorten >=1pt,auto,node distance=2cm,
  thick,main node/.style={circle,fill=blue!20,draw}]

  \node[main node] (1) {1};
  \node[main node] (2) at (0, -\ec) {2};
  \node[main node] (3) at (\hauteur*\ec, -\ec/2) {3};
  \node[main node] (23) at (\ec+\hauteur*\ec, -\ec/2) {23};
  \node[main node] (12) at (\ec+2*\hauteur*\ec, 0) {12};
  \node[main node] (13) at (\ec+2*\hauteur*\ec, -\ec) {13};
  

  \path[every node/.style={font=\sffamily\small}]
   (1) edge [draw=pink, line width=\wi mm] node [left] {} (2)
   (1) edge [draw=blue, dashed, line width=\wi mm] node [left] {} (2)
   (2) edge [draw=pink, line width=\wi mm] node [left] {} (3)
   (2) edge [draw=blue, dashed, line width=\wi mm] node [left] {} (3)
   (3) edge [draw=pink, line width=\wi mm] node [left] {} (1)
   
   (12) edge [draw=pink, line width=\wi mm] node [left] {} (23)
   (23) edge [draw=pink, line width=\wi mm] node [left] {} (13)
   (13) edge [draw=pink, line width=\wi mm] node [left] {} (12)
   
   (12) edge [draw=blue, dashed, line width=\wi mm] node [left] {} (23)
   (23) edge [draw=blue, line width=\wi mm] node [left] {} (3)
   (13) edge [draw=blue, bend left, line width=\wi mm] node [left] {} (23)
   
   
%
%

   (3) edge [draw=blue, loop above, distance=4*\ec mm, line width=\wi mm] node [left] {} (3)

;
\end{tikzpicture}
\caption{Digraph  of pairs for semi-automaton of Figure \ref{f4}.}
\label{f5}
\end{figure}

 For the next theorem, we need to define 
the digraph  of pairs. For a semi-automaton, the digraph  of pairs represents the image of each pair of states by the transition function.  An example is given  in  Figure \ref{f5}.
 The formal definition, which remains valid for digraphs  of pairs of sets of nonnegative matrices with no zero row, 
is the following.

\begin{defi}[Digraph of pairs] We call the digraph  of pairs $\mathcal{F}( M )$ the digraph  defined as follows.
\renewcommand\labelitemi{\tiny$\bullet$}
\begin{itemize}
\item One node for each unordered pair of states with repetition ($\frac{n(n+1)}{2}$ nodes),
\item an edge from pair $i_1, i_2$ to pair $j_1, j_2$ if there is a matrix $A_k \in  M $ such that $(A_k)_{i_1j_1}, (A_k)_{i_2j_2} > 0$ (or $(A_k)_{i_1j_2}, (A_k)_{i_2j_1} > 0$), where $i_1, i_2, j_1, j_2$ are not necessarily different.
\end{itemize}
\label{graph_pairs}
\end{defi}

\begin{thm}[Eppstein {\cite[Theorem 4]{eppstein90reset}}] 
$ M $ is synchronizing if and only if, in its  digraph  of pairs (as defined above), from each node representing a pair, there is a path to a node representing a single state.
\label{pairs_algo} 
\end{thm}

\begin{thm}[Eppstein {\cite[Theorem 8]{eppstein90reset}}] Deciding whether a given semi-automaton has a synchronizing word of length at most $l$ is   NP-hard.
\label{NP-hard}
\end{thm}

\section{From Consensus to Column-Primitivity}
\label{s:from_to}

We prove the equivalence between the existence of a sequence of transition matrices that drives the system to consensus and column-primitivity. Recall that column-primitivity of a set $ M $ is the existence of a positive-column  product, that is, a product of matrices from $ M $ that has a positive column. Intuitively, a positive column corresponds to an agent influencing all agents. In particular agents with the greatest and smallest values are influenced by this agent and will get closer to each other. When this happens sufficiently often, the system converges to consensus. 

Without loss of generality, we assume that the set $S$  of states is equal to  $\{1, \dots, n\}$ and the alphabet $\Sigma$ equal to $ \{1, \dots, m\}$. We note $A_w$ as  an abbreviation  for $A_{w_l} \dots A_{w_1}$.

\begin{prop}For a set $ M  = \{A_1, \dots, A_m \}$ of stochastic matrices, the answer to Problem \ref{pb21} is positive if and only if $ M $ is column-primitive.
\begin{proof}
\emph{If:} 
 Let $A_w$ have a positive  $j^{\text{th}}$  column. 
Defining $a = \min_i \{(A_w)_{ij}\}$, and using the stochasticity of $A_w$, we have
\begin{align*}\forall x \in \mathbb{R}^n, i \in  S, \;  (A_w x)_i  &= \sum_{k \in N\backslash \{j\}} (A_w)_{ik} x_k + (A_w)_{ij} x_j\\
&\leq \max_h \{ x_h \} (1 - a) + a x_j,\end{align*} and
\begin{align*}
\forall x \in \mathbb{R}^n, \; \max_i \{(A_w x)_i\} &\leq (1 - a) \max_h \{ x_h\} + a x_{j} \\
\forall x \in \mathbb{R}^n, \; \min_i \{(A_w x)_i\} &\geq (1 - a) \min_h \{ x_h\} + a x_{j}, \end{align*}
from which follows:
\begin{align*}\forall x \in \mathbb{R}^n, \; \max_i \{(A_w x)_i\} - \min_i \{(A_w x)_i\} \hspace{1cm} \\
\leq (1-a) (\max_i \{x_i\} - \min_i \{x_i\}).\end{align*}  
 Hence,  for the system
\begin{equation}x(t+1) = A_{w_{t \hspace{-.2cm} \mod l +1}} \; x(t),\label{sys_l} \end{equation}
$\lim_{s \rightarrow \infty} \max_i \{x_i(sl)\} - \min_i \{x_i(sl)\} = 0$. Finally, since \begin{align*}\forall x \in \mathbb{R}^n, \forall k \in  \Sigma, \; \max_i \{(A_k x)_i\} \leq \max_i \{x_i\} \\
\forall x \in \mathbb{R}^n, \forall k \in  \Sigma, \;  \min_i \{(A_k x)_i\} \geq \min_i \{x_i\},\end{align*} we can conclude convergence of System (\ref{sys_l}).

\emph{Only if:}
In \cite{CHJ12}, Proposition 1.b, we proved that if the answer to Problem \ref{pb21} is true, then there is a product $A_w = A_{w_l} \dots A_{w_1}$ such that $$x(t+l) = A_w x(t)$$ converges to consensus for any initial condition $x(0) = x_0$. This means that $\lim_{t \rightarrow \infty} A_w^t = \mathbf{1}y^\top$ for some $y^\top \textbf{1} = 1$. Therefore $\lim_{t \rightarrow \infty} A_w^t$ has a positive column and there is $t^*$ such that $A_w^{t^*}$ has a positive column which means that $w$ concatenated $t^*$ times is a positive-column  word. 
\end{proof}
\label{pos_sync}
\end{prop}

\begin{ex} The set 
$$ M  = \left\{A_1 = \begin{pmatrix} 0 & 1 & 0 & 0 \\
 0 & 0.8 & 0.2 & 0 \\
 0 & 0 & 1 & 0\\
 0 & 0 & 0 & 1 \end{pmatrix},
 A_2 = \begin{pmatrix} 0 & 1 & 0 & 0\\
 0 & 0 & 1 & 0\\
 0 & 0 & 0 & 1\\
 1 & 0 & 0 & 0\end{pmatrix}\right\}$$
 is column-primitive: the product 
 $$A_w = A_1A_1A_2A_2A_1 =  \begin{pmatrix}  0 & 0.2 & 0 & 0.8 \\
0 & 0.36 & 0 & 0.64 \\
0 & 1 & 0 & 0 \\
0 & 0.8 & 0.2 & 0 \end{pmatrix}$$
is positive-column. This implies the convergence of System (\ref{sys}) for  the  sequence $\sigma = \dots 1122111221$ and any initial condition. Indeed, one can check that
$$\lim_{t \rightarrow \infty} x(t)=\lim_{s \rightarrow \infty} A_w^s x_0 = \textbf{1}v^\top x_0   $$
where $ v^\top= \begin{pmatrix} 0 & 0.565\dots & 0.072\dots & 0.361\dots \end{pmatrix}$. 
\label{exlong}
\end{ex}

\begin{prop}[Equivalence between existence of a convergent trajectory and convergence with random switching] Problem \ref{pb21} is equivalent to the problem: "does System (\ref{sys}) converge with probability one when at each step, each transition matrix $A_k$ from set $ M $ is chosen with i.i.d. nonzero probability?". 
\begin{proof} "\emph{Only if}"  is evident.

\emph{If:} If the answer to Problem \ref{pb21} is positive,  then  there is a positive-column  product (Proposition \ref{pos_sync}). Under independent random switching this product  appears infinitely often  with probability one. $\max_i x_i(t) - \min_i x_i(t)$  decreases by a factor $(1 - \min_i \{(A_w)_{ij^*}\})$ each time $A_w$  appears and does not increase in between. From there, we can prove convergence as in the proof of Proposition \ref{pos_sync}.
\end{proof}
\label{eq_random}
\end{prop}

\section{Criteria based on Automata}
\label{s:results}

In this section, we extend Theorems \ref{cerny_thm} and \ref{pairs_algo} to finite sets of nonnegative matrices with no zero row,  corresponding to digraphs  with outdegree at least one.  
Moreover, we prove that a proof of the {\v C}erný Conjecture would also immediately extend to these sets. 

We define $\mathcal{R}_n$, the set of $n \times n$ matrices with nonnegative entries and no zero row. 
We proceed by associating a semi-automaton $ M' $ to any finite set $ M  \subset \mathcal{R}_n$ 
and by proving that any synchronizing word of $ M' $ can be transformed into a positive-column  word of $ M $ and vice-versa.

\begin{defi}[Digraph  associated with a word] Given a set $ M  = \{A_1, \dots, A_m\} \subset \mathcal{R}_n$ and a word $w = w_l \dots w_1$, we call \emph{digraph  associated with the word} $w$ the digraph  whose adjacency matrix is 
$$\begin{pmatrix}
0 & A_{w_l} &&& \\
&\ddots&\ddots&& \\
&&\ddots&A_{w_2}& \\
&&&\ddots&A_{w_1} \\
&&&&0\end{pmatrix}.$$ 
\label{def2}
\end{defi}

\begin{exbis}{exlong}
For set $ M $ of Example \ref{exlong}, the digraph associated with the word 11221 is depicted  in  Figure \ref{f1}.

\begin{figure}[h!]
\centering
\usetikzlibrary{arrows}
\newcommand{\ecv}{1.0} 
\newcommand{\ech}{1.4} 
\newcommand{\ba}{.7}
\newcommand{\wi}{.6}
\begin{tikzpicture}[->,shorten >=1pt,auto,node distance=2cm,
  thick,main node/.style={circle,fill=blue!20,draw}]

  \node[main node] (41) {4,5};
  \node[main node] (31) at (0, \ecv) {3,5};
  \node[main node] (21) at (0, 2*\ecv) {2,5};
  \node[main node] (11) at (0, 3*\ecv) {1,5};
  \node[main node] (42) at (\ech, 0) {4,4};
  \node[main node] (32) at (\ech, \ecv) {3,4};
  \node[main node] (22) at (\ech, 2*\ecv) {2,4};
  \node[main node] (12) at (\ech, 3*\ecv) {1,4};
  \node[main node] (43) at (2*\ech, 0) {4,3};
  \node[main node] (33) at (2*\ech, \ecv) {3,3};
  \node[main node] (23) at (2*\ech, 2*\ecv) {2,3};
  \node[main node] (13) at (2*\ech, 3*\ecv) {1,3};
  \node[main node] (44) at (3*\ech, 0) {4,2};
  \node[main node] (34) at (3*\ech, \ecv) {3,2};
  \node[main node] (24) at (3*\ech, 2*\ecv) {2,2};
  \node[main node] (14) at (3*\ech, 3*\ecv) {1,2};
  \node[main node] (45) at (4*\ech, 0) {4,1};
  \node[main node] (35) at (4*\ech, \ecv) {3,1};
  \node[main node] (25) at (4*\ech, 2*\ecv) {2,1};
  \node[main node] (15) at (4*\ech, 3*\ecv) {1,1};
  \node[main node] (46) at (5*\ech, 0) {4,0};
  \node[main node] (36) at (5*\ech, \ecv) {3,0};
  \node[main node] (26) at (5*\ech, 2*\ecv) {2,0};
  \node[main node] (16) at (5*\ech, 3*\ecv) {1,0};


   \path[every node/.style={font=\sffamily\small}]
   (13) edge [draw=pink, line width=\wi mm] node [left] {} (24)
   (23) edge [draw=pink, line width=\wi mm] node [left] {} (34)
   (33) edge [draw=pink, line width=\wi mm] node [left] {} (44)
   (43) edge [draw=pink, line width=\wi mm] node [left] {} (14)
   (14) edge [draw=pink, line width=\wi mm] node [left] {} (25)
   (24) edge [draw=pink, line width=\wi mm] node [left] {} (35)
   (34) edge [draw=pink, line width=\wi mm] node [left] {} (45)
   (44) edge [draw=pink, line width=\wi mm] node [left] {} (15)

  (11) edge [draw=blue, line width=\wi mm] node [left] {} (22)
  (21) edge [draw=blue, line width=\wi mm] node [left] {} (22)
  (21) edge [draw=blue, line width=\wi mm] node [left] {} (32)
  (31) edge [draw=blue, line width=\wi mm] node [left] {} (32)
  (41) edge [draw=blue, line width=\wi mm] node [left] {} (42)

  (12) edge [draw=blue, line width=\wi mm] node [left] {} (23)
  (22) edge [draw=blue, line width=\wi mm] node [left] {} (23)
  (22) edge [draw=blue, line width=\wi mm] node [left] {} (33)
  (32) edge [draw=blue, line width=\wi mm] node [left] {} (33)
  (42) edge [draw=blue, line width=\wi mm] node [left] {} (43)

  (15) edge [draw=blue, line width=\wi mm] node [left] {} (26)
  (25) edge [draw=blue, line width=\wi mm] node [left] {} (26)
  (25) edge [draw=blue, line width=\wi mm] node [left] {} (36)
  (35) edge [draw=blue, line width=\wi mm] node [left] {} (36)
  (45) edge [draw=blue, line width=\wi mm] node [left] {} (46)

%
%
%
   
;
\end{tikzpicture}
\caption{Digraph  associated with the word 11221 for $ M $ defined in Example \ref{exlong}. The edges corresponding to matrix $A_1$ are in black, those corresponding to matrix $A_2$ are in grey.}
\label{f1}
\end{figure}
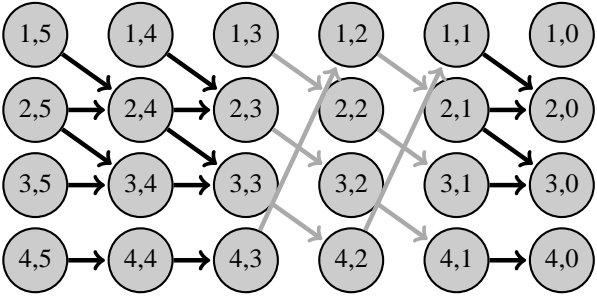

\end{exbis}

\begin{defi}[ Zero  pattern domination]
We write $A \succeq B$ and say that matrix $A$ \emph{dominates} matrix $B$ 
if  the digraph  corresponding to $A$ has more edges than that corresponding to $B$, or formally  
\begin{equation}A_{ij} = 0 \Rightarrow B_{ij}=0.\label{dominate}\end{equation} 
\end{defi}

The next definition and the proof technique of Lemma 1 are inspired by a similar construction in \cite[Theorem 17]{BJO14}.
\begin{defi}[Semi-automaton associated with a set] Let $S = \{A_1, \dots , A_m\} \subset \mathcal{R}_n$. We call the \emph{semi-automaton associated with the set $ M $}, the semi-automaton $ M' $ containing all binary stochastic matrices that are dominated by some matrix of $ M $, that is 
$$ M'   \triangleq  \{A' \;|\; A' \in \mathcal{B}_n \text{ and } \; \exists A \in  M  \text{ s. t. } A \succeq A'\;\},$$
with $\mathcal{B}_n  \triangleq  \{A \; | \; A \in \{0,1\}^{n \times n}, A\textbf{1} = \textbf{1}\}$ the set of binary stochastic matrices.
\end{defi} 
Note that the associated semi-automaton can contain a large number of matrices but we will only use it in proofs and we will not construct it explicitly in any algorithm. 

\begin{exbis}{exlong}
The semi-automaton associated with the set $ M $ of Example \ref{exlong} is $  M'  = \left\{A'_1 , A'_2,A'_3 \right\}$,
 with $$A'_1 = \begin{pmatrix} 0 & 1 & 0 & 0 \\ 0 & 0 & 1 & 0 \\ 0 & 0 & 1 & 0\\ 0 & 0 & 0 & 1 \end{pmatrix}, \; 
A'_2 = A_2, \; 
A'_3 = \begin{pmatrix} 0 & 1 & 0 & 0 \\ 0 & 1 & 0 & 0 \\ 0 & 0 & 1 & 0\\ 0 & 0 & 0 & 1 \end{pmatrix}.
$$
\end{exbis}

\begin{defi}[In-tree and spanning in-tree] We call an \emph{in-tree} a digraph  in which, for a vertex $r$ called the \emph{root} and any other vertex $v$, there is exactly one directed path from $r$ to $v$. 
For a digraph  $\mathcal{G}$, we call a \emph{spanning in-tree} an in-tree that has the same set of nodes as $\mathcal{G}$ and whose set of edges is a subset of that of $\mathcal{G}$.  
\label{def:in-tree}
\end{defi}

\begin{lem}Let $ M  = \{A_1, \dots , A_m\} \subset \mathcal{R}_n$, 
 $ M' = \{A'_1, \dots , A'_{m'}\}$ 
 its associated semi-automaton and $w= w_l \dots w_1$ be a word  over the  alphabet $ \Sigma $. The word
$w $ is positive-column  for $ M $ if and only if there is a word $w' = w'_l \dots w'_1$  over the  alphabet $\{1, \dots, m'\}$ that is synchronizing for $ M'$ and such that $$\forall i \in \{1, \dots, l\}, \; A_{w_i} \succeq A'_{w'_i}.$$
\begin{proof} 
\emph{If:} The product $A_{w} = A_{w_{l}} \dots A_{w_{1}}$ dominates $A'_{w'} = A'_{w'_{l}} \dots A'_{w'_{1}}$ because each $A_{w_i}$ dominates $A'_{w'_i}$ and domination is preserved under multiplication. In particular, if $A'_{w'}$ has a positive column, the same column is positive in $A_{w}$.

\emph{Only if:}
 We call the node $(i,j)$ with $i \in \{1, \dots , n\}, \; j \in \{0, \dots , l\}$ the node  corresponding  to the  $i^\text{th}$ row   of the $(l-j)^\text{th}$ block-row   
 of the matrix of Definition \ref{def2}. 
 This numbering is represented in Figure \ref{f1}.  Suppose that $ A_w $ has a positive  $k^{\text{th}}$ column.  
  Therefore,  in the digraph  $\mathcal{G}(w)$
associated with the word $w$, from each node $(i, l)$, there exists a path to node $(k, 0)$. 
The digraph  of these paths has a spanning in-tree rooted in $k$.
In $\mathcal{G}(w)$, for each node, there is at most one outgoing edge that belongs to the spanning in-tree. 
Therefore, some edges of $\mathcal{G}(w)$ can be removed such that the digraph  still has the same spanning in-tree and each node has exactly one outgoing edge. 
We perform the corresponding operations  on the
matrices that form the  product $A_{w_l} \dots A_{w_1}$, that is, we set to zero positive elements that do not correspond to edges of the spanning in-tree and such that on each row of each matrix, exactly one element remains positive. Then, we set to 1 all remaining positive elements.
We obtain a new product $A'_{w'_l} \dots A'_{w'_1}$ for which 
\begin{itemize}
\item the  $k^{\text{th}}$  column is positive
\item $\forall i, \; A_{w_i} \succeq A'_{w'_i}$
\item $\forall i, \; A'_{w'_i}$ is binary and stochastic  by construction,
\end{itemize}
from which we conclude that each $A'_{w'_i}$ belongs to $ M' $ the semi-automaton associated with the set $ M $ and that $w'$ is synchronizing for $ M' $.
\end{proof}
\label{extract}
\end{lem}

\begin{exbis}{exlong}

The digraph  associated with the word 11221 is represented  in  Figure \ref{f2}. The in-tree is in black. We see that removing the dashed edges allows keeping the in-tree and having exactly one outgoing edge from each node.
\begin{figure}[h!]
\centering
\usetikzlibrary{arrows}
\newcommand{\ecv}{1}
\newcommand{\ech}{1.4}
\newcommand{\ba}{.7}
\newcommand{\wi}{.6}
\begin{tikzpicture}[->,shorten >=1pt,auto,node distance=2cm,
  thick,main node/.style={circle,fill=blue!20,draw}]

  \node[main node] (41) {4,5};
  \node[main node] (31) at (0, \ecv) {3,5};
  \node[main node] (21) at (0, 2*\ecv) {2,5};
  \node[main node] (11) at (0, 3*\ecv) {1,5};
  \node[main node] (42) at (\ech, 0) {4,4};
  \node[main node] (32) at (\ech, \ecv) {3,4};
  \node[main node] (22) at (\ech, 2*\ecv) {2,4};
  \node[main node] (12) at (\ech, 3*\ecv) {1,4};
  \node[main node] (43) at (2*\ech, 0) {4,3};
  \node[main node] (33) at (2*\ech, \ecv) {3,3};
  \node[main node] (23) at (2*\ech, 2*\ecv) {2,3};
  \node[main node] (13) at (2*\ech, 3*\ecv) {1,3};
  \node[main node] (44) at (3*\ech, 0) {4,2};
  \node[main node] (34) at (3*\ech, \ecv) {3,2};
  \node[main node] (24) at (3*\ech, 2*\ecv) {2,2};
  \node[main node] (14) at (3*\ech, 3*\ecv) {1,2};
  \node[main node] (45) at (4*\ech, 0) {4,1};
  \node[main node] (35) at (4*\ech, \ecv) {3,1};
  \node[main node] (25) at (4*\ech, 2*\ecv) {2,1};
  \node[main node] (15) at (4*\ech, 3*\ecv) {1,1};
  \node[main node] (46) at (5*\ech, 0) {4,0};
  \node[main node] (36) at (5*\ech, \ecv) {3,0};
  \node[main node] (26) at (5*\ech, 2*\ecv) {2,0};
  \node[main node] (16) at (5*\ech, 3*\ecv) {1,0};


   \path[every node/.style={font=\sffamily\small}]

  (11) edge [draw=blue, line width=\wi mm] node [left] {} (22)
  (21) edge [draw=pink, dashed, line width=\wi mm] node [left] {} (22)
  (21) edge [draw=blue, line width=\wi mm] node [left] {} (32)
  (31) edge [draw=blue, line width=\wi mm] node [left] {} (32)
  (41) edge [draw=blue, line width=\wi mm] node [left] {} (42)

  (12) edge [draw=pink, line width=\wi mm] node [left] {} (23)
  (22) edge [draw=pink, dashed, line width=\wi mm] node [left] {} (23)
  (22) edge [draw=blue, line width=\wi mm] node [left] {} (33)
  (32) edge [draw=blue, line width=\wi mm] node [left] {} (33)
  (42) edge [draw=blue, line width=\wi mm] node [left] {} (43)

  (13) edge [draw=pink, line width=\wi mm] node [left] {} (24)
  (23) edge [draw=pink, line width=\wi mm] node [left] {} (34)
  (33) edge [draw=blue, line width=\wi mm] node [left] {} (44)
  (43) edge [draw=blue, line width=\wi mm] node [left] {} (14)
  (14) edge [draw=blue, line width=\wi mm] node [left] {} (25)
  (24) edge [draw=pink, line width=\wi mm] node [left] {} (35)
  (34) edge [draw=pink, line width=\wi mm] node [left] {} (45)
  (44) edge [draw=blue, line width=\wi mm] node [left] {} (15)

  (15) edge [draw=blue, line width=\wi mm] node [left] {} (26)
  (25) edge [draw=blue, line width=\wi mm] node [left] {} (26)
  (25) edge [draw=pink, dashed, line width=\wi mm] node [left] {} (36)
  (35) edge [draw=pink, line width=\wi mm] node [left] {} (36)
  (45) edge [draw=pink, line width=\wi mm] node [left] {} (46)
   
;
\end{tikzpicture}
\caption{Digraph  associated with the word 11221. In black: the in-tree. Without the dashed edges, the digraph  is that of  the  word 11223 of semi-automaton $ M' $.}
\label{f2}
\end{figure}
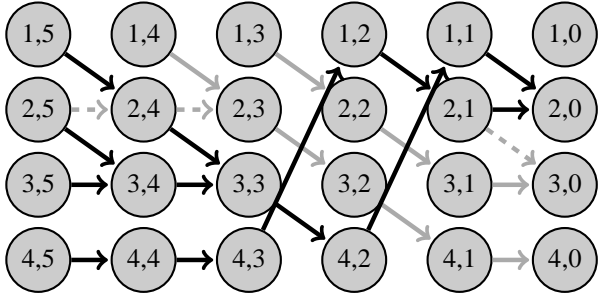
Without these dashed edges, the digraph  becomes that associated with the word 11223 of semi-automaton $ M  = \left\{A'_1 , A'_2,A'_3 \right\}$.
\end{exbis}

Lemma 1 allows extending Conjecture 1 and Theorems \ref{cerny_thm} and \ref{pairs_algo} to finite subsets of  $\mathcal{R}_n$.


\begin{thm} Let $ M   \subset \mathcal{R}_n$ be a finite set. If $ M $ is column-primitive then it has a positive-column  word of length at most $\frac{n^3-n}{6}$ and at most $(n-1)^2$ if Conjecture 1 holds. 
\begin{proof}
Suppose that $ M $ is column-primitive.
By Lemma \ref{extract}, its associated semi-automaton $ M' $  is synchronizing. Then by Theorem \ref{cerny_thm}, $ M' $ has a synchronizing word of length at most $\frac{n^3-n}{6}$. Finally, we reapply Lemma \ref{extract} to conclude that $ M $ has a positive-column  word of length at most $\frac{n^3-n}{6}$.
\end{proof}
\label{pin_stoch}
\end{thm}

\begin{thm} A finite set $ M  \subset \mathcal{R}_n$ is column-primitive if and only if, in its digraph  of pairs, from each node representing a pair there is a path to a node representing a single state. 
\begin{proof} 
The digraph  of pairs of set $ M $ and that of its associated semi-automaton $ M' $ are the same. Lemma \ref{extract} concludes.
\end{proof}
\label{pairs_algo_bis}
\end{thm}

The digraph  of pairs can be constructed in $O(m n^4)$ operations: for each pair of nodes 
 $(i_1,i_2), (j_1,j_2)$,  we add an edge if there is $k \in  \Sigma $ such that $(A_k)_{i_1j_1}, (A_k)_{i_2j_2} > 0$.
 The reachability can be checked in $O(|V| + |E|) = O(n^4)$ operations  (for example with a depth-first search algorithm). 
\begin{exbis}{exlong} On the digraph  of pairs of the set $ M $ as defined in Example \ref{exlong},
  from each pair,  there is a path to a single state (Figure \ref{f3}). Hence, $ M $ is column-primitive. 

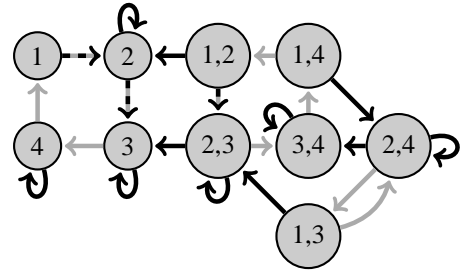
\begin{figure}[h!]
\centering
\usetikzlibrary{arrows}
\newcommand{\ec}{1.2}
\newcommand{\ba}{.7}
\newcommand{\wi}{.6}
\begin{tikzpicture}[->,shorten >=1pt,auto,node distance=2cm,
  thick,main node/.style={circle,fill=blue!20,draw}]

  \node[main node] (1) {1};
  \node[main node] (2) at (\ec, 0) {2};
  \node[main node] (3) at (\ec, -\ec) {3};
  \node[main node] (4) at (0, -\ec) {4};
  
  \node[main node] (12) at (2*\ec, 0) {1,2};
  \node[main node] (14) at (3*\ec, 0) {1,4};
  \node[main node] (23) at (2*\ec, -\ec) {2,3};
  \node[main node] (34) at (3*\ec, -\ec) {3,4};
  
  \node[main node] (13) at (3*\ec, -2*\ec) {1,3};
  \node[main node] (24) at (4*\ec, -\ec) {2,4};

  \path[every node/.style={font=\sffamily\small}]
   (1) edge [draw=pink, line width=\wi mm] node [left] {} (2)
   (1) edge [draw=blue, dashed, line width=\wi mm] node [left] {} (2)
   (2) edge [draw=pink, line width=\wi mm] node [left] {} (3)
   (2) edge [draw=blue, dashed, line width=\wi mm] node [left] {} (3)
   (3) edge [draw=pink, line width=\wi mm] node [left] {} (4)
   (4) edge [draw=pink, line width=\wi mm] node [left] {} (1)
   
   (12) edge [draw=blue, line width=\wi mm] node [left] {} (2)
   (23) edge [draw=blue, line width=\wi mm] node [left] {} (3)
   
   (14) edge [draw=pink, line width=\wi mm] node [left] {} (12)
   (34) edge [draw=pink, line width=\wi mm] node [left] {} (14)
   (23) edge [draw=pink, line width=\wi mm] node [left] {} (34)
   (12) edge [draw=pink, line width=\wi mm] node [left] {} (23)
   (12) edge [draw=blue, dashed, line width=\wi mm] node [left] {} (23)
   
   (24) edge [draw=blue, line width=\wi mm] node [left] {} (34)
   (14) edge [draw=blue, line width=\wi mm] node [left] {} (24)
   (13) edge [draw=blue, line width=\wi mm] node [left] {} (23)
   (13) edge [draw=pink, bend right, line width=\wi mm] node [left] {} (24)
   (24) edge [draw=pink, line width=\wi mm] node [left] {} (13)
   
   (34) edge [draw=blue, loop, in=150, out=120, distance=3.5*\ec mm, line width=\wi mm] node [left] {} (34)

   (2) edge [draw=blue, loop above, distance=4*\ec mm, line width=\wi mm] node [left] {} (2)
   (3) edge [draw=blue, loop below, distance=4*\ec mm, line width=\wi mm] node [left] {} (3)
   (4) edge [draw=blue, loop below, distance=4*\ec mm, line width=\wi mm] node [left] {} (4)
   (23) edge [draw=blue, loop below, distance=3.5*\ec mm, line width=\wi mm] node [left] {} (23)
   (24) edge [draw=blue, loop right, distance=4*\ec mm, line width=\wi mm] node [left] {} (24)

;
\end{tikzpicture}
\caption{Digraph  of pairs of $ M $ (Example \ref{exlong}). The black edges represent matrix $A_1$, the grey ones matrix $A_2$.}
\label{f3}
\end{figure}

\end{exbis}

\section{Shortest Positive-Column  Word: NP-hardness}
\label{ss:np-hard}

Deciding whether a semi-automaton has a synchronizing word of length at most $l$ is NP-hard (see Theorem \ref{NP-hard}). In fact, even approximating the length of the shortest synchronizing word with a fixed accuracy is NP-hard  \cite{Berlinkov-apx}.

These results extend directly to the column-primitivity problem for finite subsets of $\mathcal{R}_n$, because a semi-automaton is a particular subset of $\mathcal{R}_n$.
However, we show here that the problem remains NP-hard \emph{even if the matrices have positive diagonal elements.}
Restricting a problem can make it easier and in particular, for sets with positive diagonals, checking column-primitivity  is  easier.
$ M $ is column-primitive if and only if its digraph 
$\mathcal{G}( M )$, defined as the union of the digraphs  $\mathcal{G}(A_k)$ of which $A_k$ are the adjacency matrices, has a spanning in-tree.
In that case, there is a positive-column  word of length at most $n-1$ (given by the spanning in-tree). 


\begin{prob}
Given a set $ M \subset \mathcal{R}_n$ of matrices with positive diagonals, does there exist a positive-column  word of length smaller or equal to $l$?
\label{sw_diag}
\end{prob}

\begin{thm}
Problem \ref{sw_diag} is NP-hard. 
\begin{proof}
We proceed by reduction from 3-SAT. A 3-SAT instance consists of variables $X_1, \dots, X_v$, clauses $C_1, \dots, C_c$ of the form $L_1 \lor L_2 \lor L_3$ where each $L_i$ is a literal, that is, either a variable or the negation of a variable. The problem is to determine whether the formula $C_1 \land \dots \land C_c$ is satisfiable. 

Given a 3-SAT formula $F$ with $v$ variables, we construct a set of matrices that has a positive-column  word of length smaller or equal to $v$ (in fact never strictly smaller) if and only if the 3-SAT problem is satisfiable. 

We define a set $ M $ of $2v$ matrices. 
The matrices have size $(1+c+v) \times (1+c+v)$. 
The matrix $A_{X_i}$, representing literal $X_i$, has ones on the diagonal, a 1 in position $(1+i, 1)$ and a 1 in position $(1+v+j,1)$ for every $j$ for which assigning $X_i$ to TRUE satisfies clause $C_j$. 
The matrix $A_{\lnot X_i}$  has ones on the diagonal,  a 1 in position $(1+i, 1)$ and a 1 in position $(1+v+j,1)$ if assigning $X_i$ to FALSE satisfies clause $C_j$.

Let $P = A_{w_l} \dots A_{w_1}$ be a product of matrices from $ M $.
From the structure of the matrices, one can verify :

\emph{Claim 1: $P$ has a positive $(i,1)$ element if and only if at least one of the matrices $A_{w_k}$ has a positive $(i,1)$ element.}

%
%

We prove that the  formula  $F$ is satisfiable if and only if $M $ has a positive-column  product of length at most $v$.

\emph{If: } 
Let $P = A_{w_l} \dots A_{w_1}$ be a positive-column product of length $l\leq v$.
The first row of $P$ is equal to $\begin{pmatrix} 1 & 0 & \dots & 0\end{pmatrix}$, and therefore, the column that is positive is the first one. For each variable $X_i$ then either $A_{X_i}$ or $A_{\lnot X_i}$ appears in the product.  Indeed, if it is not the case, Claim 1 says that $P_{1+i,1} = 0$ 
and $P$ is not positive-column. 
  Also, at most one among $A_{X_i}$ and $A_{\lnot X_i}$ appear in the product. Indeed, the presence of both of them  would imply 
that $P$ is longer than $v$, contradicting our assumption.  

By Claim 1, $P_{1+v+i, 1} >0$ implies that for some $k$, $(A_{w_k})_{i1} > 0$ meaning that assigning $X_k$ to TRUE if $w_k = X_k$ and to FALSE if $w_k = \lnot X_k$ satisfies clause $C_i$.
Since $P_{1+v+i, 1} >0$ holds for any clause $i$, 
this assignment is satisfiable.

\emph{Only if:} For a satisfiable assignment, let us define  $P = A_{w_v} \dots A_{w_1}$ where $A_{w_i} = A_{X_i}$ if $X_i$ is assigned to TRUE and $A_{w_i} = A_{\lnot X_i}$ if $X_i$ is assigned to FALSE. The length of this product is clearly $v$. We prove that its first column is positive. 
The element $(1,1)$ is positive in all matrices $A_{w_k}$, therefore, it is positive in $P$. The element $(1+i,1)$ is positive in matrix $A_{w_i}$ because this matrix is equal to either $A_{X_i}$ or $A_{\lnot X_i}$. 
Finally, the element $(1+v+i,1)$ is positive in one of the matrices $A_{w_k}$ because one of these matrices correspond to assigning the variable $X_k$ to satisfy clause $C_i$.  By Claim 1, we 
conclude that the first column of $P$ is positive.

\end{proof}
\label{np-hard_thm}
\end{thm}

\begin{ex} Consider the formula $$(\lnot X_1 \lor \lnot X_2 \lor \lnot X_3) \land (X_1 \lor X_2 \lor X_3) \land (\lnot X_1 \lor X_2 \lor \lnot X_3).$$   In the reduction of Theorem \ref{np-hard_thm}, we construct the set $$ M  = \left\{A_{X_1},A_{\lnot X_1},A_{X_2}, A_{\lnot X_2},A_{X_3}, A_{\lnot X_3} \right\}$$
with
\begin{align*}
A_{w} &= \left( \begin{array}{c|c}
1&0 \\
\hline
v_w&I_{v+c \times v+c} 
\end{array} \right) \\
v_{X_1} &= \left(\begin{array}{ccc|ccc} 1&0&0&0&1&0 \end{array} \right)^\top.
\end{align*} 
The first 1 represents the assignment of variable $X_1$;  the second  1  represents satisfaction of the second clause. Similarly, the other matrices are defined with
\begin{align*}
v_{\lnot X_1} &= \left(\begin{array}{ccc|ccc} 1&0&0&1&0&1 \end{array} \right)^\top \\
v_{X_2} &= \left(\begin{array}{ccc|ccc} 0&1&0&0&1&1 \end{array} \right)^\top \\
v_{\lnot X_2} &= \left(\begin{array}{ccc|ccc} 0&1&0&1&0&0 \end{array} \right)^\top \\
v_{X_3} &= \left(\begin{array}{ccc|ccc} 0&0&1&0&1&0 \end{array} \right)^\top \\
v_{\lnot X_3} &= \left(\begin{array}{ccc|ccc} 0&0&1&1&0&1 \end{array} \right)^\top.
\end{align*}

Assignment $\lnot X_1, X_2, \lnot X_3$ corresponds to  the  product 
$$A_{\lnot X_1} A_{X_2} A_{\lnot X_3} = A_w$$
with $v_w = \left(\begin{array}{ccc|ccc} 1&1&1&2&1&3 \end{array} \right)^\top.$ 
The first column is positive because the assignment is satisfiable.

\end{ex}

\section{Concluding Remarks}

We have provided an algorithm to determine the existence of a switching sequence that makes a consensus system converge.  Our result directly applies when the switching  can be controlled, but is also relevant in case of random switching, as we have shown the equivalence between these two questions.
 Our algorithm has a better complexity than previous approaches  \cite{CHJ12} 
but we do not know if this complexity is optimal.

We have also studied the existence of  switching sequences that make the system converge, and that are periodic with a small period.
 One the one hand, we have shown that finding the  sequence that makes the system converge  with the smallest period is NP-hard. On the other hand, we have shown the existence of a  sequence that makes the system converge and that has a  small period.
 

%
%
%



\bibliography{references}{}
\bibliographystyle{plain}

\end{document}